# Observational evidence for a dry dust-wind origin of Mars seasonal dark flows

*Mathieu Vincendon, Cédric Pilorget, John Carter, Aurélien Stcherbinine*

Institut d'Astrophysique Spatiale, Université Paris-Sud, CNRS, Université Paris-Saclay, 91400 Orsay, France (mathieu.vincendon@u-psud.fr).

**Abstract**

Seasonal flows on warm slopes, or recurring slope lineae ("RSL"), have been presented as strong evidence for currently flowing water on Mars. This assumption was supported by a correlation between activity and warm temperatures, and by the spectral identification of hydrated salts. Here we first demonstrate that salts spectral identification is not robust, and that flow activity occurs on a wider range of seasons and slope orientations than previously thought, ruling out liquid water as a probable contributor. We then show that morphology, location and timing of flow activity is fully consistent with the removal and deposition of bright dust above darker underlying surfaces occurring notably in relation with seasonal dust storm activity. Mars recurring slope lineae are thus consistent with dust movements typical of present-day dry planet Mars.

## 1. Introduction

Mars is a red and black planet which has been considered with passion as a target for life searches. Changing surface features are a longstanding characteristic regularly connected to life or aqueous activity. In the 19$^{th}$ century, scientists interpreted the observation from Earth of variable dark linear features as indubitable evidence for the presence of intelligent life (Flammarion, 1892). During the 20$^{th}$ century, Mariner and Viking space observations revealed however that albedo variation on Mars primarily results from the transport of bright red dust above darker terrains (Sagan et al., 1972; Thomas et al., 1981). Later, at the end of the 20$^{th}$ century/beginning of the 21$^{th}$ century, the observation of "dark slope streaks" on warm slopes was associated with potential water flows (Ferguson and Lucchitta, 1984; Schorghofer et al., 2002). These features have since been predominantly interpreted as dry dust avalanches (Sullivan et al., 2001; Baratoux et al., 2006; McEwen 2011).

More recently, images acquired by a high-resolution camera revealed a new class of changing dark features called Recurring Slope Lineae (hereafter RSL) (McEwen et al., 2011). RSL have been presented as lengthening flows occurring each year during the warm season and/or at the warmest location, and fading during wintertime and/or at the coldest location. Liquid water-triggered flow has been argued as the most probable formation mechanism (McEwen et al., 2011, 2014; Stillman et al., 2017). The spectral identification of hydrated salts in some RSL reinforced this interpretation (Ojha et al., 2015). However, direct evidence for liquid water remains to be supplied (Ojha et al., 2015; Edwards and Piqueux, 2016), while the amount of water estimated to be necessary to form RSL may not be available (Chojnacki et al., 2016). In this context, alternative dry mechanisms have been considered and one has been shown to mimic the seasonality of one equatorial RSL site (Schmidt et al., 2017). Nevertheless, the reported seasonality of most RSL and presence of salts remain strong arguments in favour of some role for liquid water (Dundas et al., 2017), while alternative dry hypotheses should be further tested by observations.

## 2. Re-analysis of previous salt detections with near-IR spectroscopy

While salts are common on Mars and readily detected through near-IR spectroscopy (Gendrin et al., 2005), they initially remained unidentified at RSL locations (McEwen et al., 2011; Ojha et al., 2013). Spectral evidence finally published was obtained over a few spots through processing of data acquired by the CRISM imaging spectrometer (Ojha et al., 2015). Here, we reprocessed these CRISM spectral observations to assess the significance of reported detections. Ojha et al. (2015) show detections over four RSL sites (Horowitz crater, Palikir crater, Coprates Chasma, Hale crater). Detections are based on the presence of spectral bands located near 1.4, 1.9 µm, and 2.1 µm, which are interpreted as resulting from the presence of hydrated salts, in particular perchlorates, at the surface. While these three absorption features are observed together in salts analogues spectra shown by Ojha et al. (2015), most of their CRISM RSL examples contain only one (1.9 µm) or two (1.9 and either 1.4 or 2.1 µm) bands.

The most convincing hydration feature, in terms of band shape, observed in the Ojha et al. (2015) publication is the 1.9 µm band. However, this band alone is characteristic of the presence of $H_2O$ at the surface, but not necessarily salts (Carter et al., 2013). A default 2% detection threshold is generally used for the 1.9 µm band while looking for hydrated minerals in near-IR data (Carter et al., 2013). This is notably linked with the fact that this feature may be present all over the martian surface at low level due to either adsorbed water at the surface (Pommerol et al., 2009) or due to the ubiquitous presence of an OH bearing phase in surface materials (Beck et al., 2015). An averaged band depth of 1% at 1.9 µm is indeed observed in non-specific area of Mars (Beck et al., 2015), and variations depending on surface material are expected (Pommerol et al., 2009). The deepest 1.9 µm feature (2.5%) reported by Ojha et al. (2015) is located in Coprates Chasma. At that location spectra show only this 1.9 µm feature, without bands at 1.4 or 2.1 µm. We have identified the exact pixel position of the spectra shown by Ojha et al. (2015) (their supplementary material indicates "Multiple location with 1.9 absorption in the fan."). While their three detections are not aligned in the geographic projection they show (Fig. 1a), we observe in the raw, unprojected data that pixels indeed correspond to the same image column (Fig. 1b). This spectral band is in fact observed at several locations over that column, on both RSL and non-RSL sites, and can be deeper on non-RSL sites (Fig. 1c). The mapping of the presence of this 1.9 µm band with depth ≥2.5% indeed reveals that such a band appears almost exclusively over that specific column (Fig. 1b). This is a typical artefact of CRISM as the instrument collects the "vertical", along-track dimension of images through successive scans as the spacecraft progress on its orbit: all pixels of a given column in fact correspond to the same detector location. The detection of the 1.9 µm band in the CRISM observation of Coprates Chasma is thus not related to hydrated minerals at the surface, but to an instrumental artefact.

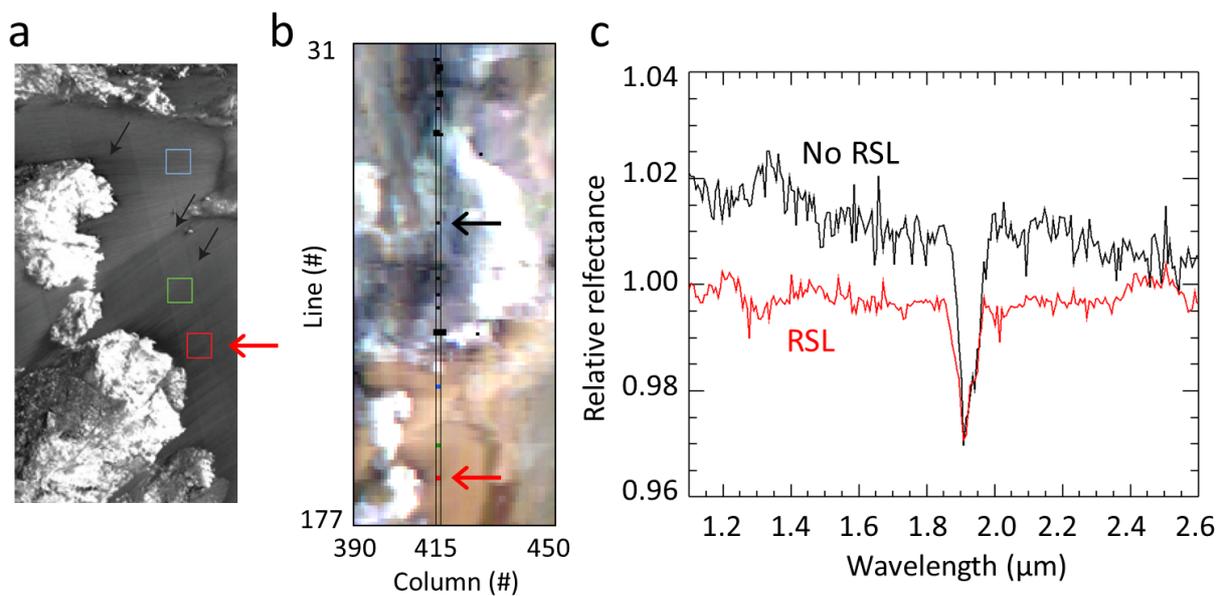

**Fig. 1.** The re-analysis of previously reported RSL hydrated minerals at Coprates Chasma show diagnostic spectral features to be related to a CRISM instrument artefact (corrupt column). **a**, HiRISE view of the RSL area containing the three CRISM "salts detection" pixels of Ojha et al. (2015) (panel adapted from Ojha et al., 2015, figure 4). **b**, Visible context of CRISM observation # FRS00028E0A in Coprates Chasma. Pixels with a hydration band (1.91 µm/1.85 µm ratio) ≥2.5% are overlaid in black, and blue/green/red for those corresponding to the three RSL "salts detection" of panel "a". Pixels belong to the same image column #415. This is a characteristic artefact of CRISM. **c**, Example of ratioed spectra showing the biased 1.9 µm band for a RSL (red) and non-RSL (black) pixel (corresponding pixels are indicated by arrows on the "b" panel, and also on the "a" panel for the red RSL pixel). Used pixels (Column #/Line #) are 415/163 over 415/162 (red) and 415/85 over 415/84 (black).

Detections at Horowitz crater show a smaller 1.9 µm band depth of 1%, but associated with a 2.1 µm feature. We have looked for the spatial distribution of these features and found again a column bias (Fig. 2.): both bands are present over a specific column. We also observe another spectral perturbation over that column at 2.47 µm that is not expected in laboratory salts analogues shown by Ojha et al. (2015). The proposed salt detection indeed correspond to a pixel of that column that present the spectral perturbation at 1.9 µm, 2.1 µm and 2.47 µm (Fig. 2). Note that this spectral perturbation results in artificial band depths that are lower or equal to 1%, which is compliant with CRISM expected performances (Murchie et al., 2007). The detections at Horowitz crater are thus also linked with instrumental systematic biases and not with minerals at the surface.

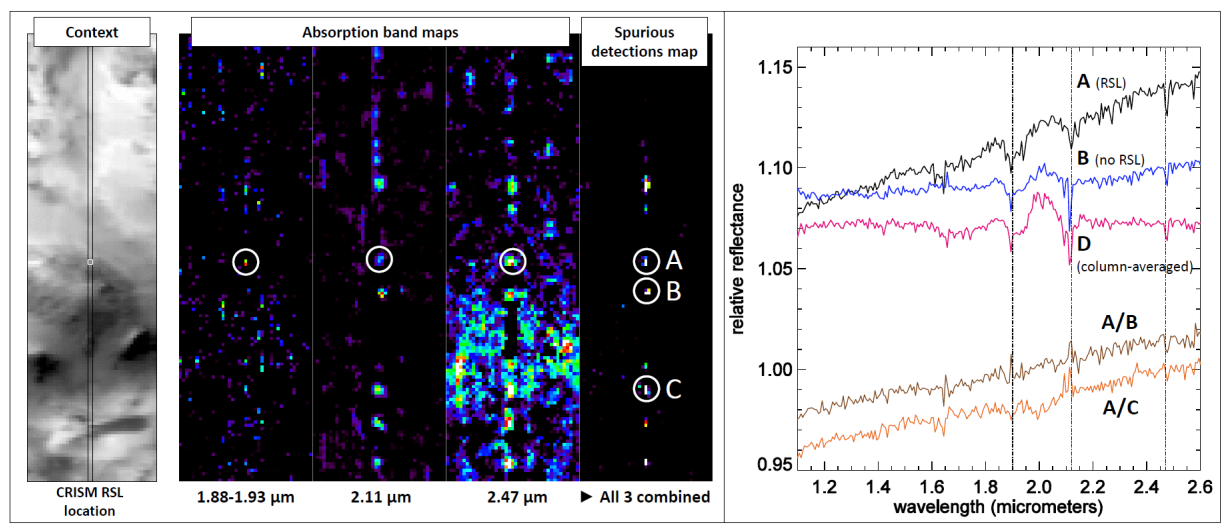

**Fig. 2.** We review data from CRISM observation # FRT00008573 (Horowitz crater) showing spectral signatures at 1.88-1.93 µm and ~2.11 µm previously interpreted as hydration bands pertaining to perchlorate salts (Ojha et al., 2015). We observe an additional sharp absorption with similar depth at 2.47 µm. A combination of the band depth maps of these 3 features revealed that the proposed detection (A) is part of a set of spurious detections located principally in the along-track (vertical) direction at the same column. Example spectra are shown on the right panel: A is the proposed RSL detection from Ojha et al. (2015), B is taken outside RSL and D correspond to the median of all spectra showing a spurious detection along the central column of the map. All clearly exhibit the same spectral features. These spurious signatures have systematic spectral noise behaviour as evidenced by the two ratios of candidate RSL spectrum A to spurious spectra B and C - see bottom spectra A/B and A/C. The proposed absorption bands disappear (i.e. amplitude equal or lower than spectral noise) in these ratios, further confirming the artefact nature of the proposed detection.

Contrary to Horowitz or Coprates, detections at Palikir and Hale are characterized by a low depth compared to the noise level and are highlighted in the Ojha et al. (2015) publication using smoothing of the original spectra (see their figures 1 and 3, compared to their figures 2 and 4). Among a large number of available pixels covering RSL, only a few ones reveal these noisy bands. Typically, we can infer from their figures that less than a few % of RSL pixels show tentative detections at Palikir and Hale craters, maybe less if we consider the large number of CRISM data investigated by the authors (Ojha et al., 2013, 2015). The large level of noise compared to band depth and the low frequency of positive pixels point toward a possible contribution of stochastic noise, in addition to the systematic effects illustrated previously, to the generation of these bands. Besides, spectra do not always show the same combination of bands among the three salts features at 1.4, 1.9 and 2.1 µm, without apparent reason for that, which is also consistent with a contribution of random noise. The spectra showing simultaneously the three bands, that should thus be the closest to salts analogues, are in fact characterized by 1.4 and 2.1 µm band shape significantly narrower compared to the laboratory examples of salts (see the Palikir green, dark blue and black spectra of figure 1c of Ojha et al., 2015). The fact that the spectral features are narrow means that only a few spectral points deviate from the continuum, which is again an indication that stochastic noise could participate to the production of some of these spectral features. We have tried to quantify this point for the '"blue" Palikir spectrum of the figure 1c of Ojha et al. (2015), one of their best examples with 3 features at 1.4, 1.9 and 2.1 µm observed in their smoothed spectrum. We show (Fig. 3) this spectrum as it appears in the data delivered to the PDS, without smoothing and without introducing optional CRISM atmospheric correction. We can see that the spectrum is very similar to the non-smoothed spectrum shown in Ojha et al. (2015). We then remove the continuum and calculate the standard deviation (σ) so as to estimate the amplitude of stochastic noise (Fig. 3). We can see that band depths are of the same order of magnitude as noise. We can see that the 2.1 µm feature, originally very narrow and difficult to identify in the Ojha et al. (2015) smoothed spectrum, is impacted in the unsmoothed data by two spectral points with anomalous values at 2.10 µm and 2.18 µm significantly ($\geq 3\sigma$) above the continuum. This systematic artefact artificially increases the ~ 2.1 µm band depth. Systematic CRISM artefacts in the 2.1 µm area have indeed been recently noticed (Leask et al., 2018). The lowest point of the 1.9 µm feature is at 1.7% from the continuum, but the depth averaged over the 6 spectral points inside the band is only 0,8%, which is not significant considering the 1% averaged band depth at 1.9 µm of Mars soils and its expected variability with soil texture or composition (see previous discussion). Finally, as the 1.4 µm feature is alternatively present or totally absent in the Palikir examples of Ojha et al. (2015), we have tried to quantify the probability for random noise to produce such a 1.4 µm band. We have performed Monte-Carlo simulations where we create a large number of flat spectra to which we add Gaussian noise. The standard deviation σ of this simulated noise is taken equal to that of the normalized data between 1.1 µm and 2.6 µm (see Fig. 3), which is an approximation as first the standard deviation of noise is expected to vary with wavelength and second data also contains systematic effects (see previous discussion). We count the number of spectra with a band with 4 contiguous points at or below ~1 σ over a wavelength range from 1.38 µm to 1.58 µm, as this range would have been considered consistent with salts (Gendrin et al., 2005; Ojha et al., 2015). 2% of the spectra meet that condition, i.e. random noise typically create a similar 1.4 µm band over 2% of the pixels. This level of expected false detection from random noise is compatible with the observed low frequency of pixels with a tentative detection of this band (see previous discussion). To summarize, the amplitude of random noise is thus too high to attribute with a sufficient level of confidence the 1.4 µm feature to the presence of salts at the surface.

Overall, our results thus show that the spectral features interpreted by Ojha et al. (2015) as evidence for hydrated salts, notably perchlorates, can be entirely explained by a combination of instrumental systematic artefacts and random noise. This result is a direct rebuttal of the statement by Ojha et al. (2015) that their "findings strongly support the hypothesis that recurring slope lineae form as a result of contemporary water activity on Mars". The presence of salts is indeed needed to reduce the freezing point of water at level compatible with RSL temperatures (Stillman et al. 2016). Consequently, the counter-detections presented in our study remove the salt argument that supported aqueous activity within RSL up to recently (Dundas et al. 2017), and argue on the contrary in favour of liquid-water free mechanisms that will be consistent with RSL observed temperatures in the absence of significant amount of salts.

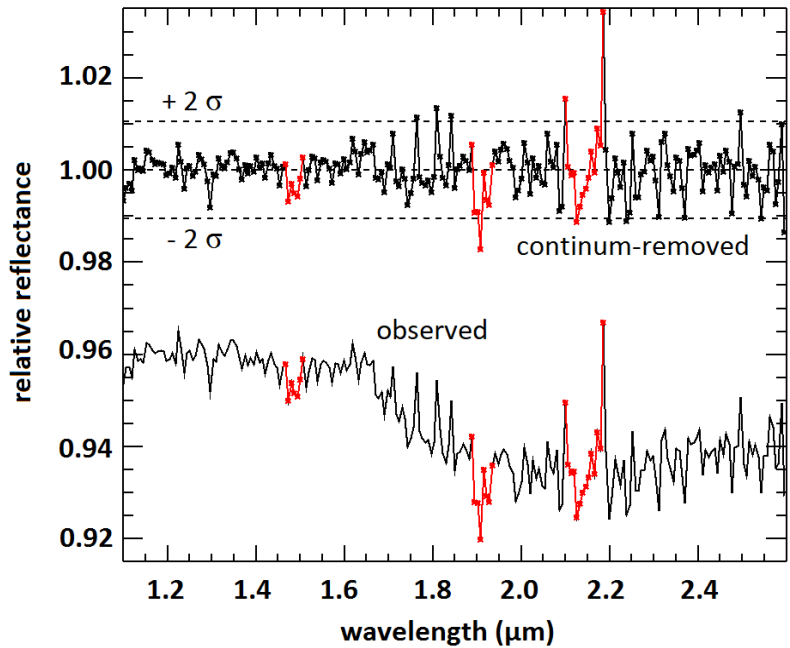

**Fig. 3.** Re-analysis of a spectrum previously interpreted as evidence for perchlorates (Ojha et al., 2015) on the basis of the spectral behaviour near 1.4, 1.9 and 2.1 µm (features highlighted in red). The standard deviation (σ) is calculated on the continuum-removed spectrum over the 1.1 – 2.6 µm range. The ±2σ level is indicated with dotted lines. The 1.9 µm band is 0.8% deep on average (see text for discussion). The 2.1 µm feature is surrounded by spectral peaks significantly (≥3σ) deviating from the average noise level, revealing a systematic artefact over that range. The band centred at 1.48 µm has deepness at ~ 1σ, a level consistent with a random noise generation (see text). The "observed" spectrum corresponds to the spectral ratio between pixel #335/218 and pixel 335/30-54 (pixels location from Ojha et al., 2015) of CRISM observations # FRT0002038F.

## 3. Re-assessment of RSL seasonality and slope orientation preference

The other major observational argument for liquid water at RSL is the correlation between RSL activity and surface temperature through seasonality and/or slope orientation (McEwen et al., 2011, 2014; Stillman et al., 2014, 2016, 2017; Dundas et al., 2017). In particular, RSL formation (apparition, lengthening) was shown to occur preferentially when or where surface temperatures are high, while RSL disappearance ("fading") was associated with decreased temperatures, which was supporting the idea of a role for $H_2O$ (McEwen et al., 2011; Ojha et al., 2014; Stillman et al. 2016). In more details, RSL formation has been initially reported at southern mid-latitudes around summer solstice on equator-facing slopes (McEwen et al., 2011). RSL have then been identified near the equator with a slope seasonality linked with solar insolation (McEwen et al., 2014; Stillman et al., 2017) and finally at northern mid-latitudes during spring (Stillman et al., 2016). A few deviant events or characteristics were also noted, like occasional RSL occurrences on pole facing slopes (Stillman et al., 2014), an overall west rather than strictly equator-facing slope preference (Stillman et al., 2017, see their figure 12), one example of early autumn lengthening (McEwen et al., 2011), and the observation of a case of "adjacent rapidly lengthening… and fading" RSL (Stillman et al., 2017; see their figure 11).

We analyse hereafter HiRISE observations of various typical RSL sites (McEwen et al., 2011, 2014; Stillman et al., 2016, 2017) to better constrain the timing and slope orientation of activity:

Lohse (43,3°S, 343.2°E), Palikir (41.6°S, 202.3°E), Raga (48.4°S, 242.4°S), Horowitz (32.1°S, 140.7°E), Rauna (35.2°N, 327.9°E) craters and Juventae Chasma (4.7°S, 298.6°E). Time of the year is indicated using Solar Longitude (hereafter $L_S$).

### 3.1. RSL activity at southern latitudes (Lohse crater)

In the southern hemisphere on the central peak of Lohse Crater, RSL activity was reported around summer solstice during the warm season. RSL were notably stated to be "active before Ls 350; didn't fade by Ls 25" (McEwen et al., 2011). HiRISE observations acquired in 2007-2008 and used in the McEwen et al. (2011) publication actually show (Fig. 4) that RSL do lengthen between $L_S$ 350° (late summer) and $L_S$ 25° (autumn), i.e. outside the $L_S$ 240° – 330° "warm summer solstice" range used for comparison with liquid water compatible temperatures (Stillman et al., 2014, 2017). Moreover, missing observations from early autumn to mid-spring prevent concluding about the presence or absence of activity in-between (Fig. 5): some RSL are in fact already present in the first mid-spring observation at $L_S$ 238° and RSL lengthening is observed up to the last observation obtained in early autumn. A recent study indeed showed that RSL activity for some southern latitude sites, for which new data with a better $L_S$ coverage have been recently acquired, is observed to start significantly earlier than previously shown, during early spring (Stillman and Grimm, 2018).

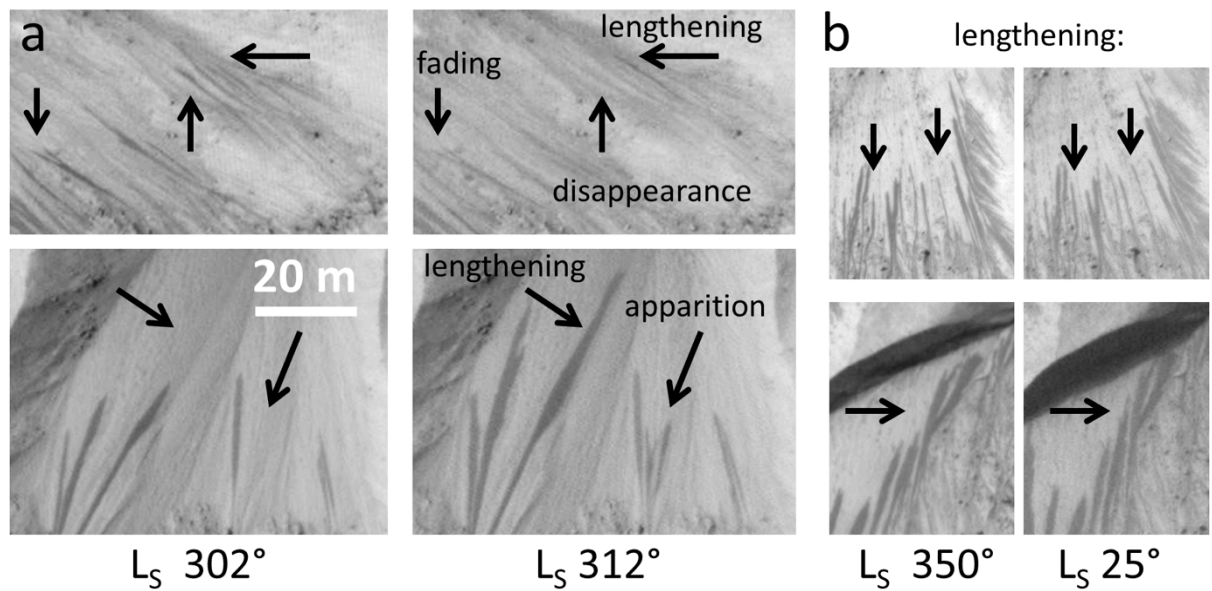

**Fig. 4.** Previously unreported RSL activity at Lohse crater (43.3°S, 343.2°E). **a**, Simultaneous RSL apparition, lengthening and disappearance in summer between $L_S$ 302° (ESP_0022697_1365) and 312° (ESP_0022908_1365) in Mars Year (hereafter MY) 30. **b**, Later summer/early autumn RSL lengthening between $L_S$ 350° MY28 (left, PSP_006162_1365) and $L_S$ 25° MY29 (right, PSP_007085_1365). North is on top.

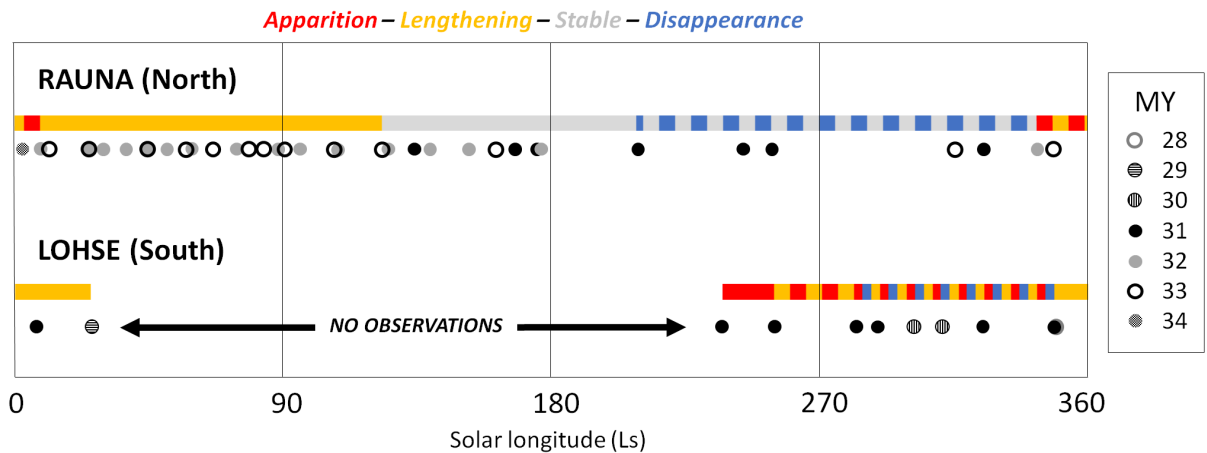

**Fig. 5.** Overview of RSL activity at Lohse crater (43.3°S, 343.2°E) and Rauna crater (35.2°N, 327.9°E). The periods over which RSL appears, lengthen, are stable or disappear are summarized for these two sites (see Fig. 4 and 6 for details). A given line goes from the first observation showing change up to the last observation showing change: actual activity periods may thus begin earlier, and end either earlier if activity actually stops prior to last observation, or later if observations are missing (Lohse). Associated uncertainties can be estimated using the timing of used HiRISE observations indicated with dots (11 for Lohse and 37 for Rauna; different MY are indicated with different symbols, see legend).

We investigated in more detail the timing of both RSL apparition or lengthening on the one hand, and fading or disappearance on the other hand. While concurrent formation and disappearance of RSL is not obvious at first glance, a detailed analysis reveals that disappearance does occur while RSL are actively forming, although with a lower amplitude (Fig. 4). Apparition and lengthening of RSL coexist with disappearance over most of southern summer at Lohse (Fig. 5). This contradicts the idea that RSL disappearance/fading occurs when temperature is insufficient to enable RSL development, as initially argued.

### 3.2. RSL activity at northern latitudes (Rauna crater)

At northern latitudes, RSL have been reported to start lengthening at $L_S \sim 340°$ and stop lengthening at $L_S \sim 150°$ (thus during late winter, spring and early summer), while subsequently fading from $L_S \sim 220°$ (mid-autumn) (Stillman et al., 2016). This timing was comparable to the rise and fall of surface temperatures but was not perfectly fitting this trend as lengthening rate was observed to be much greater in late winter/early spring (Stillman et al., 2016).

We have investigated (Fig. 5, 6) the activity for a typical northern site, Rauna crater (Stillman et al., 2016). In 2017, 37 observations over three martian years (early 31 to early 34) were available, with a variable time sampling (dense over spring/summer and sparse over autumn/fall, see Fig. 5). Observations are not available around winter solstice from $L_S$ 254° to $L_S$ 315°. RSL are first observed at $L_S$ ~345°; note however that the sparse winter sampling, with only two previous observations (over different mars years) at $L_S$ 315 and 325°, prevents a detailed analysis of the winter RSL behaviour. RSL formation or major lengthening is limited to late winter/early spring, as previously noted (Stillman et al., 2016). Between $L_S$ 25° and $L_S$ 125°, RSL only weakly lengthen. RSL are then stable up to $L_S$ 175°. At $L_S$ 210°, upper thin parts of some RSL appears faded or "blurred" (Fig. 6). Note that this $L_S$ corresponds to increased values of atmospheric dust (Kass et al., 2016) which may result in apparent contrast decrease of surface features as seen from orbit. Most parts of RSL are then clearly erased in the next available observation at $L_S$ 244°. Remaining RSL parts are stable on the two observations obtained at $L_S$ 244° and 255°. Then, additional (but not all) remaining RSL parts are erased on next available observations at $L_S$ 315°-325°. Observations gathered at $L_S$ 345° (not always coupled to a prior $L_S$ 315°-325° observation) show either a stable situation or more RSL depending on Mars years. This implies that RSL activity may start prior to $L_S$ 345° and that some RSL parts may survive all-year long, with some inter-annual variations.

These observations suggest that the formation and disappearance timing are poorly connected to surface temperatures. RSL formation is essentially occurring over a short late winter/early spring timeframe, with activity limited to gentle lengthening afterward with no new RSL and no major lengthening. Formation stops before mid-summer. Observations then suggest that RSL erasure at that location is not a uniform fading that progress from one observation to the next: erasure mainly occurs as successive events that each erases at once parts of RSL. Some parts of RSL survive almost all year long (if not all year long certain year).

### 3.3. RSL activity and slope orientation (Palikir, Horowitz and Juventae)

The prevailing slope orientation reported for southern hemisphere sites is north (equator) facing (McEwen et al. 2011), i.e. the warmest slope. At Palikir crater, a typical RSL site, RSL slope orientation is in this way referred as "North" by McEwen et al. (2011) (see their supplementary material, table S1). RSL activity however occurs on the west-facing rim, and not on the equator-facing rim. Moreover, HiRISE dense multi-coverage is restricted to this west-facing rim, while the other rims have been observed only 2 or 3 times each. As a consequence, it appears difficult to conclude to a preferential equator-facing orientation for RSL at Palikir.

Equator or near-equator facing orientation is yet a preferred orientation for many other southern latitudes sites such as Raga crater (Stillman et al., 2014). Nevertheless, temperature is not the only parameter that is correlated with slope orientation over that latitude range. A north-south slope asymmetry has been highlighted: equator-facing slopes are on average steeper than pole-facing slopes (Kreslavsky and Head, 2003). RSL are observed to be associated with steep slope (McEwen et al., 2011; Dundas et al., 2017), thus preferential activity of RSL on some equator-facing slopes could be caused by slope steepness as opposed to slope orientation. We also observe that slope morphology may not be consistent between different slopes, which may represents another factor that could impact where RSL occur. For example, at Raga crater, the density of small-scale topography such as rocks and channels is lower on the pole-facing slope.

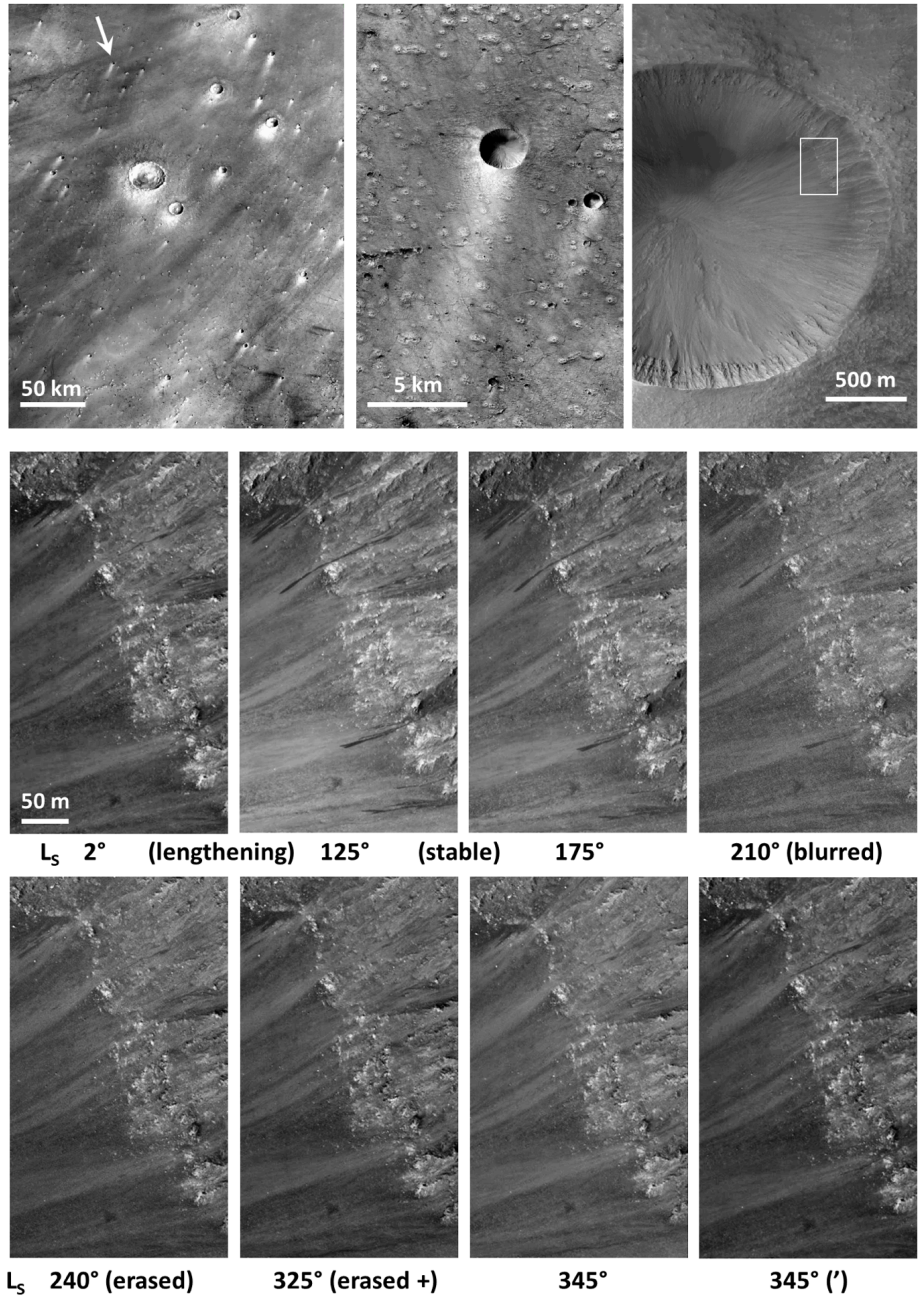

**Fig. 6.** RSL and windblown features in the area of Rauna crater (35.2°N, 327.9°E). We show on the top three panels the context area of Rauna crater (indicated by an arrow on the upper left panel), showing dark wind streaks, as already reported by Thomas et al. (2003), and bright streaks in the lee of craters. On the bottom panels (location indicated by a square on the top right panel), we illustrate the typical time sequence of RSL activity at Rauna with a selection of 8 images over a total of 37 gathered from early MY31 to early MY34. The spatial pattern of RSL is similar each year. Most upslope parts of RSL are already formed during late winter ($L_S < 360°$) and lengthen up to $L_S$ 125° with a variable rate (see text). RSL are then stable up to $L_S$ 175°. RSL appears slightly faded or blurred at $L_S$ 210°. RSL are then mainly erased in the next observation at $L_S$ 240°, and then additional (but not all) remaining RSL parts are again erased in subsequent observations at $L_S$ 315°-325°. We show two observations obtained at $L_S \approx 345°$ to illustrate the significant interannual differences observed in late northern winter. Corresponding images are: top panels, from left to right, HRSC H1619_0000_ND4 (MY27, $L_S$ 196.9°), CTX F02_036767_2154_XN_35N032W and HiRISE ESP_027840_2155; Botton panels, HiRISE ESP_50547_2155 (2°), ESP_027840_2155 (125°), ESP_028829_2155 (175°), ESP_029607_2155 (210°), ESP_030530_2155 (250°), ESP_032020_2155 (325°), ESP_050191_2155 (345°), ESP_041264_2155 (345° (')).

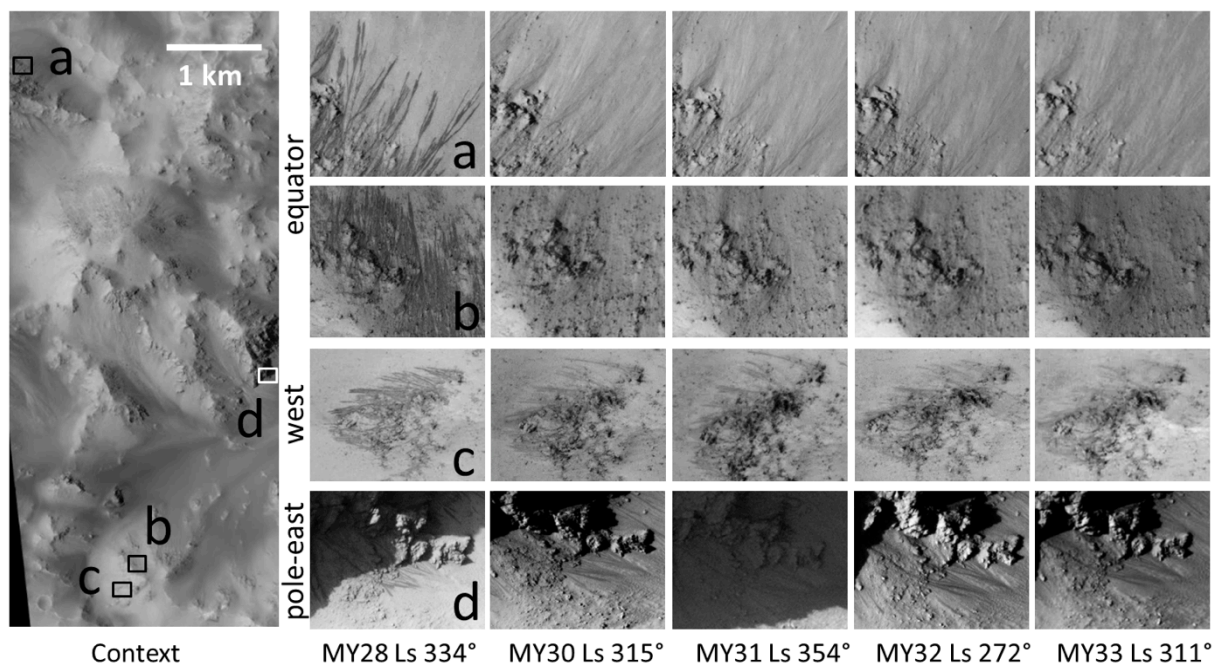

**Fig. 7.** RSL activity at Horowitz crater (32.1°S, 140.7°E) as a function of slope orientation and Mars year. Typical equator (**a**, **b**), west (**c**) and pole/east (**b**) facing slopes of mounds located inside Horowitz crater (context on left panel) are shown during summer for 5 Mars years. A global dust storm occurred during MY28 when RSL were largely observed on all slopes at $L_s$ 334°. RSL did not reappear at all on several equator facing slopes (a, c) subsequent years up to MY33, while being observed, with a lower intensity, on west (c) or east/pole-facing slopes (d). Used HiRISE images are #ESP_05757_1475, ESP_22968_1475, ESP_32726_1475, ESP_39702_1475 and ESP_49341_1475. North is on top.

On the contrary, numerous slopes on the central structure of Horowitz crater, located in the southern mid-latitudes, have similar texture and steepness (Fig. 7). Previous analyses have revealed that RSL do not show major slope preference there (McEwen et al., 2011; Stillman et al., 2014). While this behaviour was considered anomalous with respect to the majority of RSL sites, it may correspond to the standard RSL behaviour when slope have similar geomorphological properties for all slope orientations. We have analysed RSL activity for this Horowitz site (Fig. 7). Numerous RSL were observed during late summer of MY28: this post-global dust storm period is indeed known to have led to the detection of more RSL compared to other years (McEwen et al., 2011; Stillman et al., 2014). While equator-facing slopes were largely covered by RSL at that time, RSL did not reappear at all on several equator-facing slopes (Fig. 7) during the subsequent 4 Mars Years. On the other hand, RSL did reappear on some nearby slopes with other orientation such as west and pole-east facing (Fig. 7). This observation is not consistent with the idea that RSL activity is predominantly correlated to slope temperature. It also indicates that even though recurrence was considered as a necessary attribute for RSL classification (Stillman et al., 2016), some features considered as typical RSL (Fig. 7a) do not recur, as they are observed only over a single mars year and not over following four mars years.

Another case study to assess RSL behaviour with slope orientation is provided by Juventae Chasma, an equatorial site where RSL activity was reported on all slope orientations but with seasonality linked with maximum solar insolation (McEwen et al., 2014; Stillman et al., 2014, 2017).

Similarly to Horowitz, this site possesses slopes with comparable geomorphological context for all orientations (Fig. 8). It has been extensively covered with 31 available HiRISE observations at the time this study was conducted in 2017. We have investigated RSL activity over that location, looking notably for activity over periods supposed to be inactive according to Stillman et al. (2017). We did identify several instances of previously unreported activity: RSL were observe to form on north facing slopes during northern winter, and on east or south facing slopes during southern winter (Fig. 8). We show in Fig. 9 the resulting timing of RSL formation or lengthening at Juventae, as a function of slope orientation: all slopes are actually active over two periods, in summer and winter. Finally, we also observe at Juventae that RSL simultaneously appear and disappear (an example occurring on the same slope over a very short summer timeframe, between $L_S$ 337° and $L_S$ 345°, is shown on Fig. 8f).

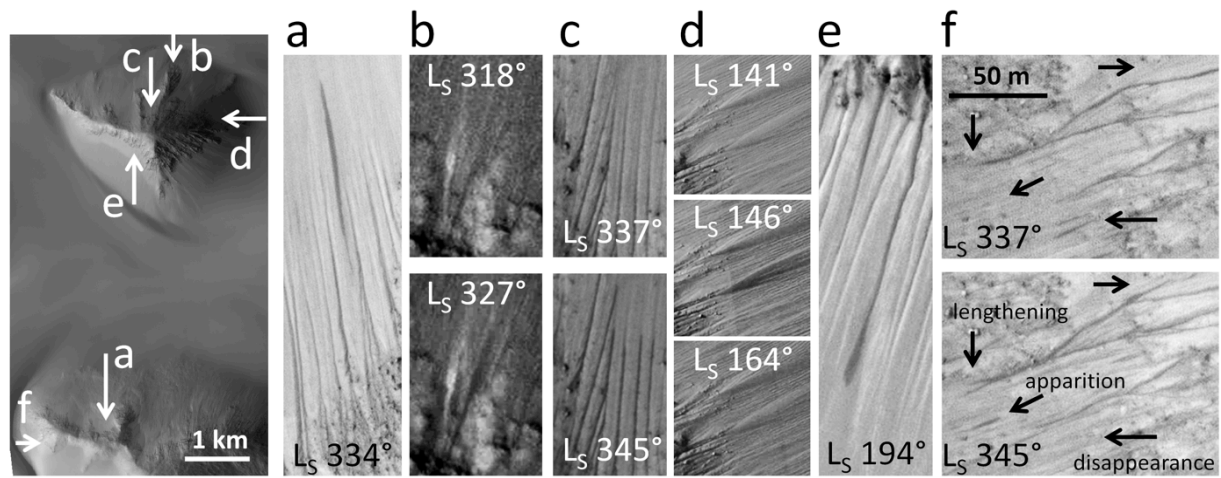

**Fig. 8.** Examples of new RSL activities reported at Juventae Chasma (4.7°S, 298.6°E). (**a**, **b**, **c**): new RSL on north facing slopes during northern winter ($L_S$ ~320 – 340°). (**d**, **e**): new RSL on east or south facing slopes during southern winter. **f**, Simultaneous rapid (<8° of $L_S$) RSL formation and disappearance during southern summer. Corresponding HiRISE images are: **context**, ESP_038509_1755; **a**, ESP_032219_1755 ($L_S$ 327° to 334°, MY31); **b**, ESP_031863_1755 and ESP_032074_1755 ($L_S$ 318° to 327°, MY31) MY31; **c**, ESP_41107_1755 and ESP_041318_1755 ($L_S$ 337° to 345°, MY32); **d**, ESP_036821_1755, ESP_036966_1755, ESP_037388_1755 ($L_S$ 141°, 146°, 164°, MY32); **e**, ESP_046909_1755 ($L_S$ 141° to 194°, MY33). **f**, ESP_041107_1755 and ESP_041318_1755. North is on top.

### 3.4. Overview of activity

We sum up in Fig. 9 the timing of RSL formation / lengthening observed in this study as a function of latitude or slope orientation (for southern latitudes we complete information gathered at Lohse – where time coverage is partial – with the recent results of Stillman and Grimm (2018) at other southern sites).

From this overview figure and previous discussions, we can summarise that: (1) All analysed sites share a common southern summer very active period, and all are also active during either southern winter or early spring (Fig. 9); (2) RSL do not show clear slope preference related to warm temperature as a function of latitude (Fig. 7) or season (Fig. 9); (3) Disappearance of RSL can be very rapid (< 10° of $L_s$) and is observed to occur contemporaneously with RSL formation/lengthening during the southern summer season (Fig. 4, 8).

These observations are poorly compatible with the idea that warm temperatures are required for RSL development, or that RSL disappearance is related to temperatures too cold to enable RSL development. These observations thus do not support a liquid water origin for RSL.

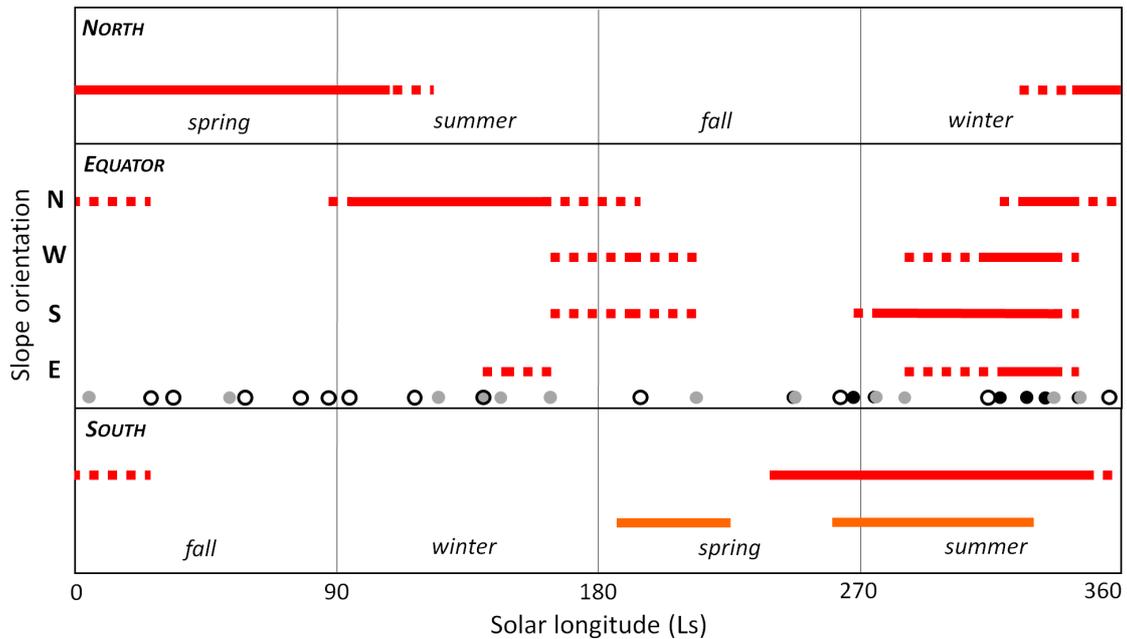

**Fig. 9.** Timing of observed RSL formation/lengthening for 3 sites (north, equator and south latitudes). In red, we show the timing of the four slope orientations (±20° azimuth) of Juventae chasma (equator), derived through the use of 31 HiRISE observations indicated with dots (see Fig. 5 for the link between symbols and MY). We also show in red the timing for Lohse (south) and Rauna (north) summarized from Fig. 5. A given line is dotted between the first or last observation showing change and the previous available observation, and straight in-between where consecutive changes are observed. In orange, the timing inferred by Stillman and Grimm (2018) at other southern sites is also shown to complete missing information at Lohse. Additional activity may remain undetected due to missing observation and observational biases.

## 4. Observed links between RSL activity and dust activity

### 4.1. Assessment of a previously proposed dry mechanism

As a contribution of liquid water to the formation of RSL is not consistent with observations, dry mechanisms should be favoured. It has been suggested that RSL may potentially form under the action of a dry gas-triggered mechanism (Schmidt et al., 2017); this mechanism indeed reproduces the seasonality of one equatorial RSL site, Garni crater (Schmidt et al., 2017). However, this mechanism is inefficient without shadows (Schmidt et al., 2017) while we regularly observed RSL formation without shadow sources (Fig. 10), in agreement with previous studies (e.g., Stillman et al., 2014). Additionally, we can notice that some RSL deviate around instead of being sourced from shadow sources like boulders (Fig. 10). Moreover, this mechanism is not efficient during dusty periods (Schmidt et al., 2017) as atmospheric dust causes light to be diffusely scattered, thus reducing shadows. While this particularity was important for this mechanism to fit the timing of activity at Garni crater, it is poorly consistent with activity peaking during the storm season (compare e.g. Fig. 9 to dust climatology by Lemmon et al., 2015 or Kass et al., 2016). It is also not consistent with the fact that RSL activity was at a maximum during MY28 (McEwen et al., 2011; Stillman et al., 2014; Fig. 7), a year characterized by a global dust storm. We can also see that the timing of activity we infer at Juventae Chasma (Fig. 9), notably the west and east-facing slopes summer activity, is not consistent with the predicted activity timing by Schmidt et al. (2017) (Juventae and Garni are very nearby sites). Lastly, this mechanism does not precisely explain one of the main specificities of RSL, that is to say the fact that RSL incrementally lengthen, as RSL growth occurs far from the putative RSL shadow source point. The main dry mechanism for RSL formation thus remains to be determined.

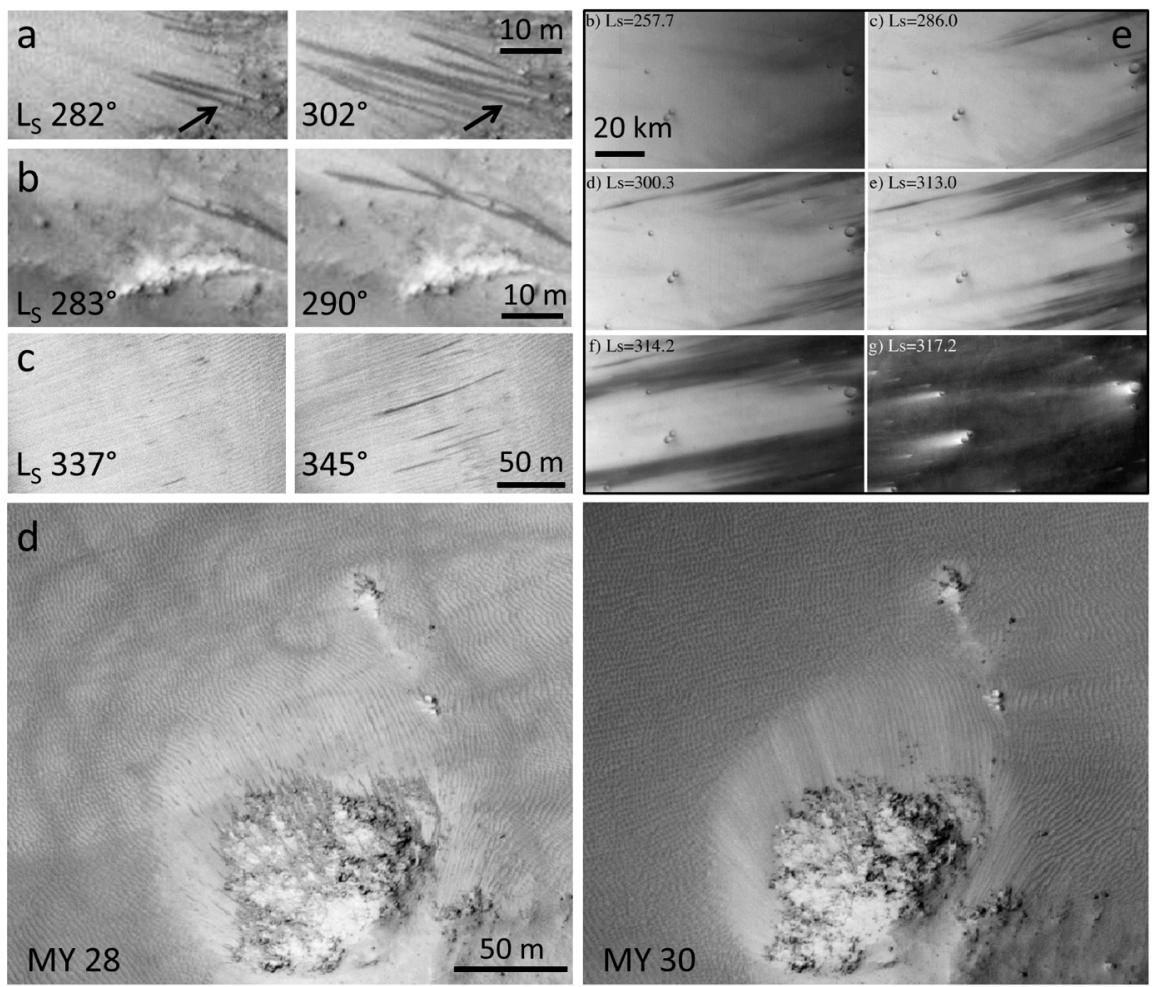

**Fig. 10.** Morphological evidence for a windblown dust origin of RSL. **a**, RSL forming a bright streak in the lead of an obstacle. **b**, RSL flow bypassing an obstacle. **c**, RSL formation without identifiable sources of dark material or shadow. **d**, RSL forming only when dust devil tracks are visible, i.e. when removable dust is available. **e**, Removal of surface dust by winds in Syrtis Major, revealing the dark underlying surface (adapted from Szwast et al., 2006, figure 18). **a**, **b**, Palikir crater. **c**, Juventae Chasma. **d**, Horowitz crater (140.78°E, 32.08°S). HiRISE observations: **a**, ESP_022267_1380 and ESP_022689_1380; **b**, ESP_039924_1380 and ESP_040069_1380; **c**, ESP_041107_1755 and ESP_041318_1755; **d**, PSP_005787_1475 and ESP_023601_1475 ($L_S$ 334° and 342° respectively); North is on top.

## 4.2. Morphological comparison between RSL and windblown dust surface features

Dark wind streaks, which form on the flat terrains of Mars, can be caused by the transport of either bright dust above darker underlying surfaces (erosional streaks), or dark sand grains above brighter surfaces (depositional streaks), and can recur every year with similar seasonal patterns (Thomas et al., 1981; Veverka et al., 1981; Rodriguez et al., 2010; Geissler et al., 2010). The lack of source material at the origin of a given class of dark wind streaks is considered as indicative of an erosional rather than depositional mechanism (Toyota et al., 2011). Similarly, RSL are observed to form in the middle of bright slopes without dark material source (Fig. 10c) and can disappear in ≤10° of $L_s$ (Fig. 4, 8): this is poorly compatible with a flow of dark grains. Such activity is, however, fully compatible with the transport of bright dust over darker underlying terrains, a thin-sheet phenomenon (Lichtenberg et al., 2007) that occurs without change of thermal inertia (Geissler 2005; Szwast et al., 2006; Toyota et al., 2011; Vincendon et al., 2015) similarly to RSL (Edwards and Piqueux, 2016; Stillman et al., 2016).

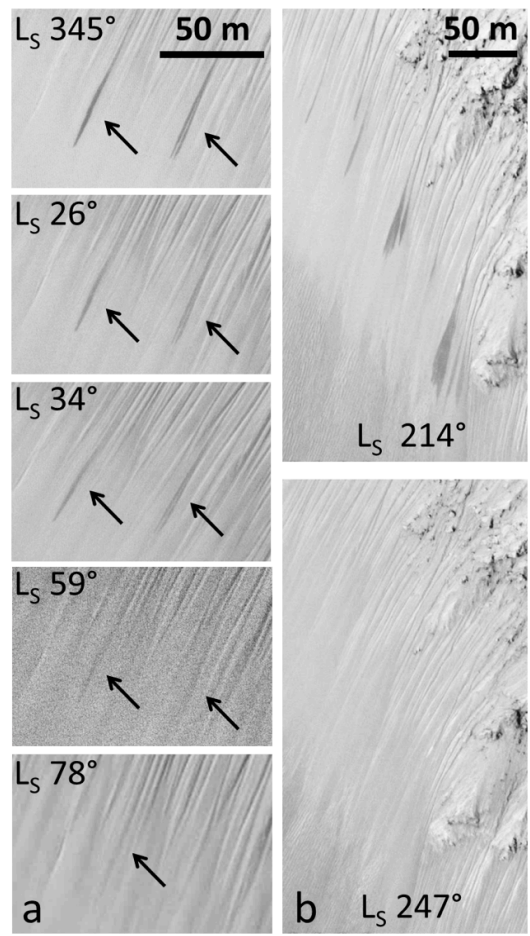

**Fig. 11.** Examples illustrating the two types of RSL disappearance (gradual and sudden) on the same location at Juventae Chasma (location "e" on the left panel of Fig. 8). **a**, RSL that disappear gradually over several images spanning a solar longitude range of 93° in late summer and autumn (MY32/33). The RSL indicated by the left arrow is still discernible at the end of this range. **b**, RSL that disappear totally at once from one observation to the next during spring of MY32 ($L_s$ gap 33°; see also Fig. 4 and 8 for smaller gaps at other sites). Used HiRISE images are ESP_41318_1755, ESP_42376_1755, ESP_42597_1755, ESP_43309_1755 and ESP_43876_1755 (panel a), and ESP_038509_1755 and ESP_039195_1755 (panel b).

Removal of bright dust by winds occurs seasonally in link with the storm season and creates wind streaks (Newman et al., 2002), features that present morphological similarities with RSL such as near-parallel dark lineae (Thomas et al., 1981, 2003; Cantor et al., 2006; Toyota et al., 2011; Fig. 6, 10, 13a) and bright streaks in the lee of obstacles (Szwast et al., 2006; Fig. 6, 10). These phenomena can result in dark features that progressively incrementally lengthen over periods of ~ 100° of LS (Szwast et al., 2006; Fig. 10e), similarly to RSL.

Dust deposition above darker underlying terrain then erase these features, either suddenly when rapid horizontal transport of bright dust occur in relation with the storm season (Szwast et al., 2006; Geissler et al., 2010, 2016; Vincendon et al., 2015) or progressively through dust settling notably during the storm season decay (Szwast et al., 2006; Geissler et al., 2010). We observe that RSL disappearance can also be either sudden or gradual: RSL disappear from one observation to the next over short $L_S$ ranges <10° during the southern spring/summer seasons (Fig. 4, 8, 11) while progressive RSL disappearance is observed in southern autumn (Fig 11), this later case being comparable to the RSL autumn fading previously reported (McEwen et al., 2011).

### 4.3. Observed proximity between RSL sites and dust deposition/lifting features

At global scale and first order, RSL are distributed (Stillman et al., 2017) close to bright /dark albedo boundaries, which correspond to transition areas between optically thick dust covered bright areas, and dust-devoid dark areas, thus to an intermediate thin dust cover (Szwast et al., 2006; Geissler, 2005; Geissler et al., 2016; Vincendon et al., 2015). This spatial correlation is thus consistent with RSL being formed through removal of a thin dust cover above a dark substrate.

Removal of bright dust by dust devils results in dark tracks when the underlying substrate is darker than dust (Cantor et al., 2006; Greeley et al., 2006; Lichtenberg, et al. 2007). At Horowitz crater, where RSL formation is restricted to the global dust storm year on certain slopes (Fig. 7), we observe that adjacent dust devil tracks formation is also restricted to the same year (Fig. 10d), suggesting that RSL formation occurs at that location only when removable dust is available at the surface.

In the northern hemisphere, we also observe a spatial correlation between RSL and the existence of an underlying dark substrate above which bright dust motions are observed. We highlight in Fig. 12 transient bright slope streaks observed over this dark substrate, and the coexisting adjacent RSL on the same crater wall. The albedo of dark RSL and bright surroundings is comparable to that of the nearby dark substrate and bright streaks respectively (Fig. 12). These observations are consistent with the presence of a dark underlying substrate all over that crater slope, and show that bright dust can be mobilized to form transient streaks above that substrate. Note also that both bright streaks and RSL are changing at same $L_S$ (Fig. 12), which is consistent with a common activation process, similarly to dark and bright streaks that may occur in similar wind conditions (Thomas et al., 2003).

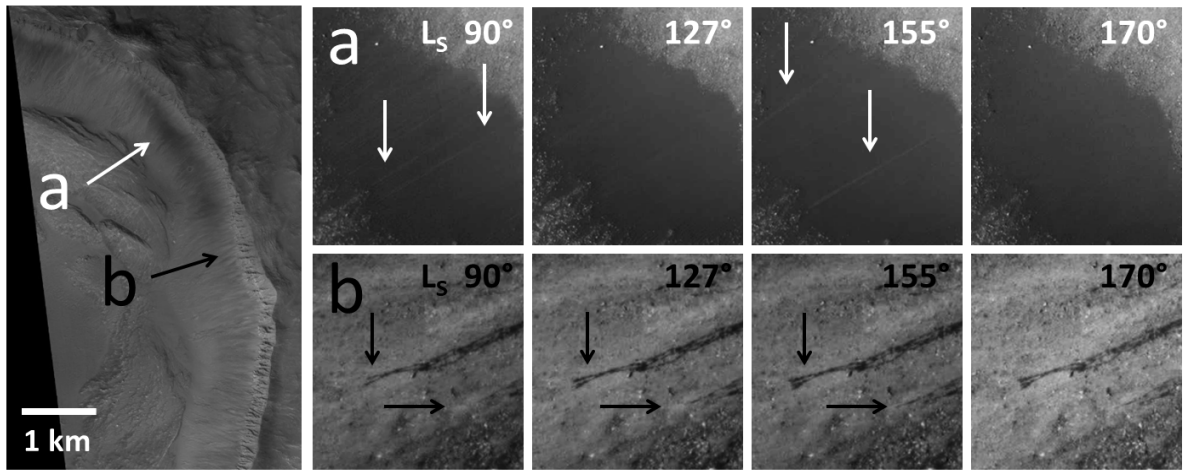

**Fig. 12.** Transient bright linear streaks and slow RSL lengthening co-occurring on the same crater wall located at 37.2°N, 320.4°E. **a**, bright streaks are observed above a dark substrate; streaks appear and disappear from one observation to the next. **b**, RSL are active on the same crater slope at the same summer solar longitudes. HiRISE images are ESP_033919_2175 (context), ESP_35409_2175, ESP_36464_2175, ESP_37176_2175, and ESP_37532_2175 (subsets). North is on top.

Similarly, recent studies have highlighted the spatial proximity between RSL and other dark features on slopes such as dark bands (Schaefer et al., 2017), slope streaks (Stillman et al., 2017) or slumps (Chojnacki et al., 2016), which may also indicate the presence of an underlying dark substrate at these RSL slopes.

We observe the co-occurrence of RSL and wind streaks at Rauna crater (Fig. 6). As reported by Thomas et al. (2003), the area of Rauna crater corresponds to one of the few location where mesoscale linear streaks have been reported (see their figure 2d). We indeed observe in that area several features (large-scale wind streaks, bright streaks in the lee of craters, Fig. 6) previously interpreted as resulting from the removal or deposition of bright dust above a darker substrate linked with wind activity. This strongly suggests the presence of both a dark substrate and wind activity able to mobilize dust at Rauna crater.

We have investigated (Fig. 13) another crater (199.7°E, 31.1°N) also located in area covered by major mesoscale wind streaks according to Thomas et al. (2003), and near Propontis where surface albedo changes have been observed (Szwast et al., 2006; Geissler, 2005; Vincendon et al., 2015). Moderate spatial resolution data reveal dark slope streaks within this crater, i.e. large-scale slope flow generally interpreted as dry mass wasting of thick dust deposits revealing either a dark substrate, or a bright dust layer with a different surface state (Sullivan et al., 2001; Baratoux et al., 2006; Chuang et al., 2010; Brusnikin et al., 2016). Nearby wind streaks and albedo change features suggest the presence of a relatively thin dust cover in that area, which may indicate that these slope streaks correspond to the exposure of a darker substrate. Note that these slope streaks occur slightly north/west of previously reported slope streaks (Aharonson et al., 2003; Schorghofer et al., 2007). At high resolution we observe small-scale features with size and morphology that may present similarities with some northern hemisphere RSL (Fig. 13). The potential seasonality of these small-scale features cannot be assessed (only one HiRISE observation is available). We note that the larger-scale slope streaks appear fresh in northern spring observations, and faded or erased in winter observations (Fig. 13), a timing compatible with northern hemisphere RSL activity. Future observations with increased time and spatial sampling may be of interest at that location, to further investigate the links between RSL, slope streaks and wind streaks.

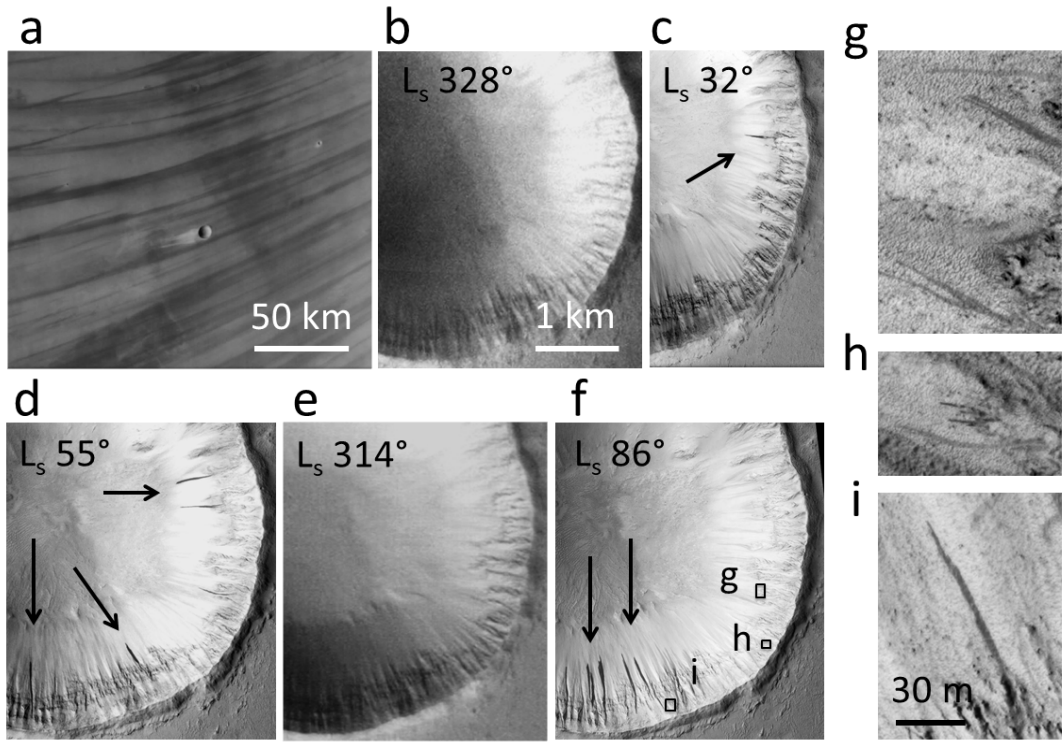

**Fig. 13.** Properties of dark slope flows observed in an area covered by linear wind streaks. **a**, crater at 199.7°E, 31.1°N located within previously reported dark linear wind streaks (Thomas et al., 2003). **b to f**, large-scale features (arrows) corresponding to slope streaks (Sullivan et al., 2001) are observed on crater walls. The use of five different images obtained between 1999 and 2014 provide some partial information about the timing of these slope streaks: these images show fresh dark flows at $L_s$ 32° to 86° (**c,d, f**) and no fresh flow at $L_s$ 314-328°. **g to i**, small-scale dark slope flows that present some size and morphological similarities with RSL are also observed on these slopes, but only once due to missing hi-resolution imagery coverage (localisation are indicated on panel **f**). **a**, MOC #0101483 (1999, $L_s$ 138°). **b**, THEMIS #V1746702 (2005, $L_s$ 328°). **c**, MOC #S1602427 (2006, $L_s$ 32°). **d**, CTXB18_016755_2140 (2010, $L_s$ 55°). **e**, HRSC #H9528 (2011, $L_s$ 314°). **f,g,h,i** HIRISE #035308_2115 (2014, $L_s$ 86°). North is on top.

### 4.4. Seasonal timing comparison between RSL and surface-atmosphere dust exchanges

Overall, we finally observe a general timing correlation between RSL activity (both apparition and disappearance) and the seasonal cycle of surface – atmosphere exchange of dust.

During autumn and early winter ($L_s$ ~ 0° – 140°), the content of atmospheric dust decreases to its lowest values (Lemmon et al., 2015). Outside polar caps edge, dust storm activity is reduced at that time (Cantor et al., 2001; Kass et al., 2016). Surface lifting inferred to be caused by dust devil activity is principally observed in the northern hemisphere (Cantor et al., 2006). For RSL, this period also corresponds to both limited activity and activity favouring the northern hemisphere: gentle lengthening for the most active northern sites (Stillman et al., 2016; Fig. 6) and modest activity compared to other $L_S$ at some equatorial sites (McEwen et al., 2014; Schmidt et al., 2017). Fading compatible with progressive dust settling is also observed at that time, at equatorial or southern latitudes (Fig. 11; McEwen et al. 2011).

After $L_S$ ~140°, the optical depth of dust increases, first at southern and equatorial latitudes (Smith 2004), as a result of an increased dust lifting activity at low to mid-latitudes (Guzewich et al., 2015; Lemmon et al., 2015). During that late southern winter/early spring period, RSL form on all slopes of the Juventae equatorial site (Fig. 9) and start a small pulse of lengthening at southern latitudes (Stillman and Grimm, 2018).

The main maximum of the storm season (for usual, non-global dust storm years) occurs between $L_S$ 220° and ~240° according to dust optical depth monitoring (Cantor et al., 2001; Smith 2004; Lemmon et al., 2015; Kass et al., 2016). At that time rapid horizontal transport of bright dust have been reported to erase dark terrains (Szwast et al., 2006; Vincendon et al., 2015) while sudden erasure of RSL occur (Fig. 6, Fig. 11).

Optical depth then decreases up to $L_S$ ~270 – 300° (Lemmon et al., 2015), and dust storm activity is reduced (Guzewich et al., 2015), suggesting deposition of dust at the surface and a relative decrease in the efficiency of dust lifting at that time. This period indeed corresponds to limited RSL formation (Fig. 9) and RSL disappearance at northern and equatorial latitudes.

Dust optical depth and dust storm activity then increases again up to a second maximum at $L_S$ ~310°- 330° (Guzewich et al., 2015; Lemmon et al., 2015). The formation of erosional wind streaks has been reported at that time (Thomas and Veverka, 1979; Veverka et al., 1981; Szwast et al., 2006; Fig. 10e), as well as surface albedo changes (Szwast et al., 2006). This period immediately follows dust deposition and corresponds to increased dust lifting activity; it also corresponds to major episodes of RSL formation and lengthening at all latitudes and slope orientation (Fig. 9), together with some cases of rapid disappearance (Fig. 4, Fig. 8).

While main trends of the dust seasonal cycle are relatively consistent each year (Lemmon et al., 2015) and homogenous over low to mid-latitudes (Vincendon et al., 2009), annual and spatial variations are also observed (Cantor et al., 2001), the most flagrant example being global dust storms. The fact that RSL formation is favoured during years characterised by a global dust storm (e.g., Fig. 7) is fully consistent with RSL being formed by dust movements: large amounts of unstable dust can be deposited in certain area of Mars after global dust storms (Szwast et al., 2006; Vincendon et al., 2015), and this dust is indeed progressively lifted, revealing the dark substrate, over a time range compatible with RSL activity (Szwast et al. 2006; Fig. 10e). Similarly, the lower intensity and modified timing of northern hemisphere RSL (Fig. 9) is consistent with the observed spatial distribution of regional dust storms and dust optical depth as a function of time of the year favouring the southern hemisphere at low to mid-latitudes (see e.g. Smith 2004; Kass et al., 2016). It is also consistent with north/south differences in dust lifting efficiency that can be inferred from the timing of dust devil tracks formation (Balme et al., 2003; Fisher et al., 2005; Whelley and Greeley, 2006; Cantor et al., 2006).

## 5. Discussion and conclusion

In this paper, we have reviewed the observational evidence previously derived from visible imagery and near-IR spectral imagery in favour of a contribution of liquid water to the initiation and/or development of RSL still considered in recent studies (Dundas et al., 2017). We showed in section 2 that instrumental biases have been misinterpreted as evidence for salts at RSL. We then showed in section 3 that the link between RSL activity and warm season or warm slope orientation is only partial and that there are other factors that may influence where/when RSL form. Finally we showed in section 4 how the spatial, morphological and timing properties of RSL are consistent with the removal and deposition of bright dust above darker underlying terrains.

The observation of concurrent RSL lengthening and rapid disappearance is consistent with dust removal and horizontal dust transport occurring simultaneously when winds are strong enough to mobilise dust. The progressive nature of dust removal for most RSL is consistent with the progressive, wind-induced increase of wind streaks, further suggesting that winds may contribute to the removal of dust that results in RSL. Wind-induced surface material movements are observed to be seasonal near the equator (Ayoub et al., 2014): activity is peaking during the southern summer period, and is still present but at lower level in southern winter, a timing that present some similarities with RSL activity (section 4.4).

Slope winds are an additional local component to regional winds and are expected to exceed regional winds on certain slopes (Spiga and Lewis, 2010), which may explain, in addition to the reduced grains stability on steep slopes (Dundas et al., 2017), why RSL are usually observed in area lacking flat terrain wind streaks. Slope winds are sensitive to temperature gradient and forcing by solar insolation (Thomas et al., 1981; Savijarvi and Siili, 1993; Magalhaes and Gierasch, 1982), similarly to local turbulence (Whelley and Greeley, 2006; Fenton and Michaels, 2010), which may explain the partial correlation between RSL and solar insolation. The contribution of winds influenced by heating and thermal instability has thereby been previously suggested for the triggering of dark slope streaks (Sullivan et al., 2001).

We observe that several RSL follow every year similar paths corresponding to small-scale topography and frequently spread from channel exits (Fig. 6, Fig. 8, Fig. 10). Channels are expected to increase wind speed through Venturi effect (Chrust et al., 2013). Winds may thus more easily reach dust-lifting threshold (e.g. through saltation of underlining dark sand grains) within channels and at channel endings.

Dust removal may in some case propagate downslope in the form of a dry avalanche, and explain bright fans observed at the end of some RSL (Ojha et al., 2013; Chojnacki et al., 2016) and the links between RSL and granular flows angles (Dundas et al., 2017). This may particularly apply to RSL that form at once and do not incrementally lengthen, as such behaviour is comparable to slope streaks interpreted as dry avalanches (Sullivan et al., 2001; Baratoux et al., 2006). In this interpretation, RSL and slope streaks formation mechanisms may share some common properties (Fig. 13), and differences may be notably related to the thickness of the dust cover.

Overall, the RSL phenomenon is thus fully consistent with a dry aeolian process where activity results from the seasonal exposure and masking of dark surfaces through surface-atmosphere exchange of dust.

**Acknowledgments**

The first author would like to thank the persons who have supported us during the review process, in particular Enora, and also Aymeric (between first submission to another journal in September 2017 and acceptance in February 2019, the paper was evaluated five times by six referees, and remained with reviewers or editors over 94% of that ~ 500 days period).